\documentclass[twocolumn]{article}

\usepackage{arxiv}

\usepackage[utf8]{inputenc} 
\usepackage[T1]{fontenc}    
\usepackage{hyperref}       
\usepackage{xurl}           
\usepackage{booktabs}       
\usepackage{amsfonts}       
\usepackage{nicefrac}       
\usepackage{microtype}      
\usepackage{graphicx}
\usepackage{natbib}
\usepackage{doi}
\usepackage{longtable}      
\usepackage{array}
\usepackage{enumitem}       

\newenvironment{cellitemize}
  {\leavevmode\vspace{-\baselineskip}%
   \begin{itemize}[nosep,leftmargin=*,label=\textbullet,
                   topsep=0pt,partopsep=0pt,parsep=0pt,itemsep=0pt]}
  {\end{itemize}}

\title{Assessing Company Contributions\\
to Societal Resilience\\
\large Extending the Societal Capacity Assessment Framework to Agentic AI}

\date{}    

\author{Catherine Simons \\
	Cambridge Boston Alignment Initiative\\
	\texttt{cesimons@mit.edu} \\
	\And
	Alexander K.~Saeri \\
	MIT FutureTech\\
	\texttt{aksaeri@mit.edu} \\
	\And
	Peter Slattery \\
	MIT FutureTech\\
	\texttt{pslat@mit.edu} \\
	\And
	Neil Thompson \\
	MIT FutureTech\\
	\texttt{neil\_t@mit.edu} \\
}

\renewcommand{\headeright}{}
\renewcommand{\undertitle}{}
\renewcommand{\shorttitle}{Assessing Company Contributions to Societal Resilience}

\hypersetup{
pdftitle={Assessing Company Contributions to Societal Resilience: Extending the Societal Capacity Assessment Framework to Agentic AI},
pdfsubject={Agentic AI, Societal Resilience, AI Governance},
pdfauthor={Catherine Simons, Alexander K.~Saeri, Peter Slattery, Neil Thompson},
pdfkeywords={Societal resilience, Agentic AI, AI governance, Corporate governance},
}

\begin{document}

\twocolumn[{%
\maketitle

\begin{abstract}
Companies that deploy AI agents and make them available to others are creating the sociotechnical circumstances under which this technology integrates into existing social and economic structures. AI-deploying companies are institutional actors that actively shape society's capacity to withstand and govern the consequences of agentic AI. In view of these societal impacts, companies can build societal resilience by designing and promoting safer implementations of AI agents. To operationalize this goal, this paper adapts the indicator-based Societal Capacity Assessment Framework (SCAF) to measure how a company's deployment decisions contribute to societal resilience, inverting its original measurement of societal resilience as a backdrop for deployment decisions \citep{gandhi2025societal}. Our procedure has two steps: a conceptual step in which we design a suite of indicators that define what SCAF's vulnerability, coping, and adaptive capacities mean when assessing a company's agentic AI deployment decisions; and a measurement step in which we apply this framework in a structured assessment of public-facing Microsoft documents.
\end{abstract}

\vspace{1.5em}
}]

\section{Introduction}
AI systems are embedded in broader social contexts involving their users and operators. AI risks are shaped not only by technical model capabilities, but also by how systems are deployed and the conditions of the institutions and communities exposed to them. Feedback loops between deployment decisions and norms of use lead to societal impacts, since downstream actors live under sociotechnical circumstances that are not of their making but powerfully shape what they do. Rather than accept these impacts as a pre-determined result of technological innovation, we theorize about safer diffusion, applying the foundational Science and Technology Studies claim, ``by far the greatest latitude of choice exists the very first time a particular instrument, system or technique is introduced'' to the widespread introduction of advanced AI agents \citep{winner1980artifacts}.

As AI systems are introduced, reducing the harm from AI risks involves both reducing the likelihood of the risk materializing and increasing the resilience of society to absorb the negative impacts when a risk does materialize. AI safety scholarship tends to focus on the former problem and on how the actions of frontier AI developers create and mitigate risks \citep{fli2025index, saferai2025ratings}. We take a different approach, arguing that the companies that deploy AI agents, without necessarily training foundation models, play a significant role in shaping the societal impacts of AI.

Throughout this paper, we use \emph{AI agent} to refer to a system configured to iteratively plan, reason, and use tools to accomplish real-world tasks, and \emph{agentic AI} to refer to the broader class of products and platforms built around such systems. By \emph{societal resilience}, we mean the capacity of a social system to absorb the immediate impact of an adverse shock, adapt to a changing risk landscape to reduce the risk of shocks occurring in the first place, and transform access to resources in the face of present and future shocks -- a conceptualization we inherit from the social-ecological tradition \citep{keck2013social}.

Companies that deploy agentic AI make a range of decisions about which use cases to encourage, what level of AI autonomy is the default, and what safeguards to include. These decisions have downstream effects on third-party risk exposure, cultural norms, and institutional functioning, and these dynamics are orthogonal to how well the company measures its own risk exposure. Tech platforms increasingly mediate human life. The transition towards AI agents that act on people's behalf thus constitutes a cross-domain societal threat.

Agentic AI deployers and societal resilience are both underexplored focus areas in AI governance. One exception is the Societal Capacity Assessment Framework (SCAF), which bridges the gap between resilience studies and AI governance and offers an indicator-based method for assessing a social unit's vulnerability, coping capacity, and adaptive capacity in the face of advanced AI risk \citep{gandhi2025societal}. However, the initial prototype was designed to assess countries, and no comparable method exists for treating the company as the unit of analysis, despite the potential for companies' actions to influence societal resilience.

To fill this gap, we extend the SCAF to companies, using agentic AI as the risk domain of focus. Large companies, such as Microsoft, frame agentic products as augmenting human agency, economic productivity, and scientific innovation. However, AI safety experts caution that agentic AI may pose loss of control risks, expand cybersecurity threat surfaces, and undermine recognizable chains of human accountability in institutional decision-making \citep{bengio2026international}. Companies that deploy AI agents increase societal vulnerability to agentic AI risks. However, governments are not the only actors that can govern these risks; the same companies can simultaneously build societal resilience through design decisions that enable other actors to respond better to failure.

This motivates our core research question: how can we assess company contributions to societal resilience? We answer this by prototyping a suite of resilience indicators grounded in high-priority agentic-AI risk-management actions and applying them to Microsoft's public documents in an illustrative case study.

\section{Background and Motivation}
Existing AI governance leaves deployers who most directly govern how agentic AI reaches society comparatively unexamined. Our extension of SCAF offers a theoretically grounded assessment methodology to target this niche. Future comparative applications of our assessment can inform policy design that incentivizes socially beneficial corporate AI governance.

\subsection{Societal Resilience}
Recently, AI governance work has begun to apply the interdisciplinary tradition of ``societal resilience'' to the study of AI risks, emphasizing the need to improve society's response to harms that are likely to materialize, rather than solely focusing on upstream interventions that prevent risks from materializing \citep{bernardi2024societal}. The 2026 International AI Safety Report's final chapter on risk management concludes with a section on ``building societal resilience,'' addressing the research topic as an important area for further investigation \citep{bengio2026international}. \citet{bengio2026international} say that proactively building resilience can create an ecosystem for safe and beneficial adoption of AI technology and limit the emergence of risk from interactions with resources, individuals, organizations, and technologies. They argue that benefits of societal-resilience-building are highly diffuse, so individual stakeholders tend to underinvest in it. Policymakers are currently deciding whether and how to incentivize, fund, develop, and evaluate resilience-building measures.

The inclusion of ``societal resilience'' as a concept of interest to AI governance is laudable, but we must guard against it becoming an empty buzzword. Policy scholars have used the word to support various theoretical approaches, and tech companies have adopted it in promotional materials about their philanthropic initiatives and anti-regulation stances \citep{chandler2014resilience, hutson2024societal, openai2026industrial}. A possible objection to this usage is that it reinforces narratives about technological inevitability and passes responsibility onto downstream actors who are then cajoled to ``adjust'' to become resilient.

In contrast, the academic usage of ``resilience'' originated in ecology to refer to whether a system can absorb change and maintain relationships between components when continually confronted by the unexpected \citep{holling1973resilience}. The rise of ``complexity thinking'' cast failure as the primary medium through which systems reveal themselves and made ``failing better'' a governing aspiration \citep{chandler2014resilience}. Social scientists adapted the concept to describe how living communities cope after immediate disruption, adapt based on accumulated experience, and transform institutions to enhance access to resources in present and future crises \citep{keck2013social, bene2012resilience, folke2010resilience, walker2004resilience}. In this paper, we use the term ``resilience'' in its specific social-scientific sense.

We extend the Societal Capacity Assessment Framework (SCAF) developed by \citet{gandhi2025societal} in order to build upon their work defining a concrete indicator-based method that adapts societal resilience literature to AI safety. They proposed a framework consisting of indicators that measure a social actor's vulnerability, coping capacity, and adaptive capacity in response to AI-related risks and suggested that this can ground organizations' risk management in insights about societal conditions\footnote{Country-level analysis is demonstrated in their prototype SCAF, but they explain that this framework is designed to structure assessments of ``social units'' in general, rather than being limited to countries.}. They assessed Australia as an illustrative case study.

We used this work as a template, borrowing their design decisions about how to construct resilience capacities and structure a pragmatic, standardizable, and evidence-based approach. Rather than use the framework to measure societal resilience as a backdrop for organizations' AI risk management, we adapt it to measure company contributions to societal resilience.

\subsection{The Company as an AI Governance Actor}
Within the field of AI governance, most efforts are focused on the actions of frontier AI companies. The Future of Life Institute's AI Safety Index and SaferAI's risk-management ratings grade frontier labs against expert-defined criteria \citep{fli2025index, saferai2025ratings}. Stein-Perlman's AI Lab Watch tracks companies' safety practices and the Institute for AI Policy and Strategy compares responsible-scaling policies \citep{steinperlman2024ailabwatch, andersonsamways2024responsible}. However, the companies that modify AI agents, set up multi-agent systems, and provision agentic solutions constitute a major and neglected class of ecosystem actors. They bridge the gap between AI development and real-world diffusion.

One reason for this near-exclusive focus on frontier AI companies is that training frontier LLMs is compute-intensive, and the highly concentrated supply chain introduces choke points for government intervention. Detectability, excludability, and quantifiability make AI compute uniquely governable \citep{sastry2024computing}. Within this vein of thinking, compute access could be leveraged to demand desirable safety practices -- only companies that comply with hypothetical requirements would be allowed to conduct massive training runs\footnote{Similar reasoning pervades geopolitical analyses of AI governance -- cutting off China's compute supply chain so that only American companies can train the largest LLMs.}. In contrast, agentic AI governance is relatively diffuse, since the auxiliary scaffolding, platforms, and customization that turn a model into a deployed agent do not require centralized training runs. As such, decentralized deployers, rather than a handful of leading developers, are determining how agentic AI risk reaches society.

Conceptual work supports focusing on AI governance at companies downstream of the frontier model developers and assuming that they can contribute to societal resilience. Technocratic solutions are ill-suited for ``wicked problems'' that carry high degrees of complexity, uncertainty, and divergent stakeholder perspectives\footnote{Rittel and Webber set out ten distinguishing properties that separate ``wicked problems'' -- no definitive formulation, no stopping rule, solutions are good-or-bad instead of true-or-false, no ultimate test of a solution, every solution is a ``one-shot operation,'' no enumerable set of potential solutions, every wicked problem is essentially unique, every wicked problem is a symptom of another problem, discrepancies can be explained in numerous ways, and the planner has no right to be wrong \citep{rittel1973dilemmas}.} \citep{rittel1973dilemmas}. The ultimate goal of the AI safety and governance fields -- ``making AI go well'' -- is one such wicked problem. Technocratic intervention could look like government enforcement of proven technical standards within frontier AI companies.

Building resilience is suited for ``wicked problems'' because it doesn't depend on enumerating failures, pre-testing solutions, and overcoming definitional problems; instead, social actors are enabled to absorb, adapt to, and recover from surprises that weren't specified. These actors maintain agency over their own goals.

Our version of SCAF is designed to support decentralized AI governance -- we assume that many companies can build societal resilience, independent of whatever centralized governance is applied to frontier AI companies. Researchers have applied complexity theory to the AI governance tradeoffs between centralization and adaptability \citep{ilcic2025artificial}. They caution that decentralized markets chronically overproduce socially and environmentally harmful goods by privileging narrow valuation metrics of shareholder primacy and competitive advantage, but argue that AI should be treated as a complex adaptive system whose instability and unpredictability demand the adaptability of decentralization, rather than conventional governance. They argue that decentralized systems are more adaptable because actors have better access to information to respond quickly to local failures, and on a societal level this creates better feedback mechanisms.

This intellectual tradition informs our assumption that AI diffusion should prioritize societal resilience. Other work corroborates this emphasis on diffusion and the important role of downstream companies in shaping the trajectory of AI. A theme across recent major enterprise reports is that organizations struggle to scale agents because of governance and organizational constraints rather than technological limits \citep{bcg2026radar, mckinsey2026state, deloitte2026state}. This supports the view that the societal effects of AI are not technologically determined. The company as a governance actor can reduce others' exposure to AI risks, and improve their ability to respond to known and unknown threats.

\section{Methods}
Throughout this project, we strove to remain faithful to the original SCAF construction whenever possible. We completed a two-step process: first, we designed a company-level version of SCAF that measures contributions to societal resilience against agentic AI risk; second, we applied this company-level SCAF in an illustrative case study.

\subsection{Designing New SCAF Indicators}

\subsubsection{Resilience Capacity Definitions}
The original SCAF resilience capacity descriptions included examples tailored for country-level analysis. Their vulnerability description included ``population health'', coping included ``emergency response services,'' and adaptive included ``public policies.''

However, the core definitions were applicable to our use case, and are provided in Table~\ref{tab:capacities}.

\begin{table*}[t]
	\caption{SCAF Resilience Capacities \citep{gandhi2025societal}}
	\centering
	\begin{tabular}{p{0.18\linewidth}p{0.72\linewidth}}
		\toprule
		Capacity & Description \\
		\midrule
		Vulnerability & Structural and background conditions that increase a society's susceptibility to the risk in question. \\
		\addlinespace
		Coping & Resources and systems in place to absorb, respond to, and recover from an adverse shock in its immediate aftermath. \\
		\addlinespace
		Adaptive & Interventions aimed at mitigating, planning for, and adjusting to risks over the medium- to long-term, including measures to reduce the likelihood of adverse impacts and their severity.\newline\newline \emph{Transformative capacity:} Adaptive interventions that reshape societal systems in lasting and systemic ways. \\
		\bottomrule
	\end{tabular}
	\label{tab:capacities}
\end{table*}

We sought to operationalize these definitions in order to measure dynamic contributions to resilience capacities rather than snapshots of a society's current state. This required us to theorize about how companies can contribute to each of these capacities.

\subsubsection{Capacity Descriptions for Company Contributions}
Since the country-level examples in the original SCAF do not have direct analogues for company-level analysis, we needed to translate these resilience capacity descriptions and identify general company actions that can contribute to each capacity (Table~\ref{tab:contributions}).

\begin{table*}[t]
	\caption{Company Contributions to Resilience Capacities}
	\centering
	\begin{tabular}{p{0.18\linewidth}p{0.72\linewidth}}
		\toprule
		Capacity & Company Contribution Description \\
		\midrule
		Vulnerability & Companies contribute to vulnerability through the deployment reach of their products because diffusion of a technology increases societal exposure to risks associated with that technology. Customer use patterns and technological dependence become background conditions for future failures. When a technology is integrated into critical infrastructure, society becomes more susceptible to the risk of that technology failing. \\
		\addlinespace
		Coping & Companies can contribute to the coping capacity by offering users features that enhance the ability to immediately respond and recover from a failure. This could include technical provisions, such as monitoring tools that allow users to detect failures as soon as they require intervention. This could also take the form of customer support that assists with incident response. \\
		\addlinespace
		Adaptive & Companies can contribute to the adaptive capacity by offering users features that reduce the risk of technical failures occurring in the first place by improving users' access to information and their ability to alter agents' operation. This could include acceptable use policies that limit deployment in critical domains, as well as user guidance that promotes human-in-the-loop implementation norms. This could also include aggressive red teaming to proactively identify failure modes, as well as transparent information reporting that increases social actors' ability to learn from cumulative experience and avoid replicating others' failures.\newline\newline Companies can contribute to the transformative capacity by provisioning innovative technologies that offset the offense-defense balance in favor of defense. For example, a tool that enables the proactive discovery and effective mitigation of cyberattack vulnerabilities can have a lasting positive impact. Tools that democratize access to institutions can also transform society. \\
		\bottomrule
	\end{tabular}
	\label{tab:contributions}
\end{table*}

\subsubsection{Selection of Agentic AI Risk Category}
We needed a category that falls directly under the purview of large companies and is particularly consequential for societal impacts, such that company-level risk mitigations could translate into resilience capacities. Agentic AI risks met these criteria. In order to define measurable indicators for each of these categories of resilience contributions, we narrowed our focus to agentic AI risks.

\subsubsection{Agentic AI Threat Model}
Societal resilience is a relational concept. It doesn't exist in isolation, but rather as resilience \emph{to} something---resilience to a given threat. Our ``agentic AI risk'' category spans the near- and medium-term agentic AI risks that are expected to emerge when AI systems can act on behalf of users. Agency exists on a spectrum that spans degrees of access and autonomy, and the threat model associated with highly agentic systems differssubstantially from that associated with generative AI chatbots that are used in single-turn call-and-response scenarios.

AI systems that can do things humans do, but massively faster and in parallel, will affect the broader social ecosystem as they diffuse throughout institutions. Company-level management can help mitigate risks of associated socially harmful outcomes. Social and technological dynamics associated with autonomous general-purpose AI systems form known threat pathways that demand resilience. These concerns do not constitute an exhaustive list of agentic AI threats relevant for discussion about societal resilience, but rather an illustrative collection of adverse societal impacts that could be ameliorated via company-level resilience building.

Removing human labor bottlenecks is a core concern associated with agentic-AI-enabled automation. This threat is often discussed in the context of technological unemployment, but it is causally linked to other problems that would persist in a world where the AI job loss hypothesis were disproven\footnote{Currently there is a lack of consensus among economists about the labor market impacts of AI.}. Human labor bottlenecks make it more difficult for actors to scale projects that may maliciously or unintentionally cause harm; eliminating them provides uplift to bad ideas. A group of humans is very different from a swarm of AI agents -- humans bring different background knowledge, experience, and personalities that add friction to group operations. This friction helps people surface problems and correct course. Without design decisions that prioritize human agency, agentic AI ``groupthink'' can also undermine interrelational foundations of institutions and bring well-intentioned but harmful projects to completion.

Agentic AI introduces difficult technical security challenges. Multi-agent interactions can introduce unpredictability that undermines an evaluation-based governance paradigm, since unknown threats may emerge as a result of post-deployment interactions, even if a network's constituent agents' capabilities are well-understood. Tool use and interaction in varied contexts also introduce new opportunities for exploitation. Prompt injections are an architectural problem for LLM agents. Attackers can embed malicious instructions in any data source an agent processes and the agent may follow those commands, since agents may fail to distinguish between the user's request and content received from external sources \citep{owasp2026state}.

\subsubsection{Agentic AI Relevance to SCAF}
Agentic AI risks are a relevant domain of focus for the Societal Capacity Assessment Framework because of their potential societal impacts.

The original SCAF draws two illustrative risks from the International AI Safety Report and frontier safety frameworks: cyber and chemical and biological risks, instead of the concerns about agentic AI described above. Although their prototype assessment exclusively focuses on these domains, they are presented as examples, rather than an exhaustive implementation of SCAF. In fact, Gandhi et al.\ considered an ``autonomy and potential loss of control'' risk category, but cautioned that ``definitions that enable societal capacity assessment remain underdeveloped'' \citep{gandhi2025societal}.

When we shifted the SCAF unit of analysis to companies instead of countries, and the substance of analysis to contributions to societal resilience instead of societal conditions, we found that agentic AI definitions were sufficient to enable assessment. Agentic AI risk management falls directly under the purview of large companies, and existing work in this domain operationalizes definitions associated with ``agency'' and ``autonomy'' for pragmatic mitigation recommendations.

\subsubsection{Selecting High-Priority Agentic AI Risk Mitigations}
As in the original SCAF, where each indicator is associated with multiple data sources, each indicator in our version is associated with several risk mitigations that serve as empirical proxies for the resilience constructs (vulnerability, coping capacity, adaptive capacity). Since the mitigation landscape for agentic AI is effectively unbounded, we needed a principled, non-arbitrary selection rule for which actions our indicator suite should be designed to capture.

We used the expert-prioritized recommendations in the 2026 CLTC Agentic AI Risk Management Profile \citep{madkour2026agentic}. We selected this report because it was scoped to our agentic AI risk domain and large company unit of analysis. They selected high-priority subcategories of mitigations based on whether they were high-priority in the sibling general-purpose AI profile, or required additional consideration beyond it. The contents were grounded in the NIST AI RMF and synthesized from a wider literature.

\subsubsection{Designing Indicators for Public-facing Documents}
We sought to balance the competing aims of designing a framework applicable to public-facing documents, while also not omitting categories whose lack of disclosure would itself be a notable finding. In order to build some intuition about the kinds of practices that could be measured in these documents, we conducted an agentic-AI keyword search across a corpus of corporate AI governance documents from 201 companies and then manually reviewed documents from the six companies with the most keyword matches.\footnote{Agentic AI keyword list and match results available in supplemental materials.} This gave us general evidence about which practices could in principle be measured. We reviewed document collections from multiple companies, rather than only the company that would become our illustrative case study. We wanted to guard against overfitting the indicator design to the particular information that Microsoft happens to disclose.

\subsubsection{Incorporating Expert Prioritization}
In order to design a suite of indicators that are both consistent with the high-priority practices in the Agentic AI Risk Management Profile \citet{madkour2026agentic} and actually fit with the conceptual domains from SCAF, we classified these practices according to the resilience capacities and then defined indicators based on these lists. Categorizing each action into an associated SCAF capacity involves asking, ``Does this action primarily affect background vulnerability, harm recovery, or harm prevention?'' Once we had lists of risk mitigations for each category, we defined indicators. We noticed that the vulnerability and transformative capacities were undersupplied, and attributed this to the risk-management focus in the . We added two indicators to correct for this weakness (deployment criticality; AI-enabled defense), and then defined the remaining fourteen indicators to serve as `buckets' for the practices associated with each capacity. Reflecting the Agentic AI Risk Management Profile, our indicator distribution was still adaptive-dominant. We accepted this as a feature of the risk mitigation landscape, but future innovation in technical AI governance could create new recommended mitigations that contribute to the coping or transformative capacities.

Replicating the original SCAF design, we opted for a minimal set of indicators within each capacity. Envisioning future comparative use cases, a smaller suite of indicators improves coding reliability and company profile legibility.

\subsubsection{Prototype Company-level Agentic AI Societal Capacity Assessment Framework}
This process yielded the company-level agentic AI Societal Capacity Assessment Framework presented in Table~\ref{tab:prototype}.\footnote{See supplemental materials for original mapping against the CLTC recommendations.}

\begin{table*}[p]
	\caption{Prototype Company-level Agentic AI Societal Capacity Assessment Framework with Indicators}
	\label{tab:prototype}
	\centering
	\small
	\begin{tabular}{p{0.22\linewidth}p{0.68\linewidth}}
	\toprule
	Classes of Activities & Measurement Proxies \\
	\midrule
	\multicolumn{2}{l}{\textbf{Vulnerability}} \\
	\midrule
	1.1 Deployment Reach & Reach and criticality of the company's agent deployment across enterprise, government, and security workflows \\
	\addlinespace
	1.2 Autonomy Scope & Scope of autonomy and authority granted to agents (range of actions, tools, and decisions an agent may execute without further human approval) \\
	\addlinespace
	1.3 Anthropomorphic Framing & Organizational framing of agents through language, role assignment, and interface design that treats agents as worker- or employee-equivalent rather than as tools operating under human direction \\
	\addlinespace
	1.4 Use Boundaries & Extent to which the company has declared and bounded its agents' permitted uses, prohibited activities, and operational constraints \\
	\midrule
	\multicolumn{2}{l}{\textbf{Coping Capacity}} \\
	\midrule
	2.1 Rapid Intervention & Intervention infrastructure that halts, suspends, or contains agent actions during deployment to limit the duration and consequence of unintended or harmful behavior. \\
	\addlinespace
	2.2 Behavioral Logging & Behavioral logging and audit systems that capture and make reviewable agent interactions, decisions, and outputs throughout deployment, to detect deviations from expected behavior and support post-hoc accountability. \\
	\addlinespace
	2.3 Tiered Oversight & Risk-tiered escalation routing that directs high-risk agent decisions to human reviewers rather than to automated execution, prioritizing human attention. \\
	\addlinespace
	2.4 Error Remediation & Error-correction and remediation procedures that reverse, repair, or compensate for harms caused by agent errors or unintended actions, to restore affected parties and return systems to their expected state. \\
	\midrule
	\multicolumn{2}{l}{\textbf{Adaptive Capacity}} \\
	\midrule
	3.1 Identity Binding & Agent identity and permission controls that bind, authenticate, and constrain each agent to a verifiable identity with scoped permissions \\
	\addlinespace
	3.2 Incident Disclosure & Public incident reporting practices that disclose agent-related incidents and near-misses to industry-wide or regulator-maintained registries \\
	\addlinespace
	3.3 Discovery Bounty & External incentive programs that reward third parties for discovering and disclosing safety, security, or alignment issues in deployed agents \\
	\addlinespace
	3.4 Risk Thresholds & Pre-committed risk-tier thresholds that classify agent capabilities into measurable tiers below stated red lines \\
	\addlinespace
	3.5 Capability Evaluation & Test, benchmark, and document agent capabilities, limitations, and safety-relevant behaviors before deployment \\
	\addlinespace
	3.6 External Red Teaming & Third-party adversarial assessment that engages organizations outside the company to elicit harmful, unintended, or unauthorized behavior from deployed agents \\
	\midrule
	\multicolumn{2}{l}{\textbf{\emph{Transformative Capacity}}} \\
	\midrule
	3.7 Defensive Provision & Defensive AI capability provision through which the company develops, deploys, or shares AI-based tools strengthening cybersecurity, threat intelligence, civic integrity, or public-sector resilience beyond its own products. \\
	\addlinespace
	3.8 Customer Governance & Customer-administered governance tools that enable downstream deployers to configure and customize agents operating within their own environments. \\
	\bottomrule
	\end{tabular}
\end{table*}

\subsection{Microsoft Case Study}
To demonstrate the application of this assessment, we applied our company-level agentic AI Societal Capacity Assessment Framework to Microsoft as an illustrative case study. Rather than serve as a ``score'' for Microsoft's contributions to societal resilience, the case study is intended to show how public documents can be assessed using our extension of SCAF.

\subsubsection{Selection of Microsoft as a Case Study}
Microsoft was one of the top companies for ``agentic AI''-associated keyword matches in our initial review of organizational documents. Other candidates included Salesforce, SAP, Prosus, and Visa. Since our adaptation of SCAF was designed to be portable across companies deploying AI agents, rather than limited to a specific industry (e.g., enterprise management software, finance), other companies could have been equally valuable examples.

We use Microsoft in this case study because it is a particularly high-reach example of our assessment's intended unit of analysis. Microsoft ships agentic AI products at enterprise and government scale, and this is a particularly clear example of the systemic stakes that make societal resilience a suitable lens for analysis. Since our manual review of companies' documents found that Microsoft has unusually rich public-facing AI governance documents, we reasoned that it is also a hard case for finding gaps. This makes the gaps that our SCAF assessment surfaces more telling, and more indicative of the methodological value of our approach.

\subsubsection{Selection of Microsoft Documents}
Before we could apply our SCAF assessment, we needed to collect relevant public-facing documents. To do this systematically, we manually reviewed all of the Microsoft AI governance documents included in the initial agentic AI keyword search stage. Then, we searched for recent Microsoft publications about agentic AI.

Documents were then included according to the criterion that they represent Microsoft's own position. In practice this meant excluding standalone research papers and think-piece blog posts that do not speak to the company's practice. We also excluded older (2023) documents that discuss mitigations in the context of generative AI, since relevant actions from those documents reappear in subsequent agentic AI documents. We included a small number of research outputs (e.g., Magentic-One) where they are direct evidence of deployed capabilities.

This document review process yielded a collection of 21 documents spanning security-product blogs, formal AI governance documents, and strategy and adoption-framing material.

\subsubsection{Keyword Search for Evidence}
To systematically assess these documents' coverage of our resilience indicators, we designed keyword sets for each of our 16 indicators. Then, to find evidence for each indicator, we conducted a keyword search across the Microsoft document corpus. We iteratively refined the keyword sets, removing items that produced systematic false matches (such as ``principal'' matching ``principal researcher''). When deciding which keywords to include, we assigned a ``mentions'' count to each keyword so that no keyword with genuine matches was inadvertently dropped. We didn't mark keywords for exclusion solely because they returned zero matches. In several cases the absence of any mention of a concept across the corpus is itself a substantive finding about what the company does and does not disclose. The keyword lists are intended to be portable across company corpora, providing the backbone for a future comparative analysis of companies using our version of SCAF.

This led to 306 keywords associated with the 16 indicators\footnote{See Supplemental Materials.}. We manually reviewed all Microsoft excerpts associated with these keywords, and marked them for inclusion based on whether the text excerpt qualitatively matched an indicator. We included 142 excerpts, and coded them according to our SCAF indicators. The top keyword-matched indicators were 1.2 Autonomy Scope (23 matches), 2.2 Behavioral Logging (19 matches), 2.4 Error Remediation (15 matches), and 3.4 Risk Thresholds (25 matches).

\subsubsection{Analysis of Keyword-matched Evidence}
To identify key findings for each of the 16 indicators, we conducted a qualitative analysis of the indicator-coded text excerpts\footnote{See Appendix~\ref{app:a}.}. We summarized implications from the supporting evidence for each indicator. For example: whether disclosed practices are company commitments or aspirational language, whether responsibility is retained by the company or passed to downstream customers, and whether high-stakes capabilities are merely acknowledged rather than firmly constrained.

\section{Results \& Discussion}

\subsection{Main Findings}
This paper extended the Societal Capacity Assessment Framework \citep{gandhi2025societal} from its original country-level unit of analysis to the level of the company, and applied the resulting prototype to Microsoft, using agentic AI as the risk domain. This demonstrated the possibility of measuring contributions to societal resilience via a structured assessment of stated organizational activities. The SCAF structure bridges the gap between resilience literature and AI governance, and our agentic AI application addresses an understudied niche: corporate agentic AI governance.

\subsection{Microsoft Findings}
Detailed results from this exercise and expanded discussion are presented in Appendix~\ref{app:a}. In summary, Microsoft's SCAF profile is adaptive-capacity-dominant. This is unsurprising, since organizational risk management is essentially the process of building company-level adaptive capacity to support adjustment to a changing risk landscape, and companies have incentives to heavily invest in inward-facing risk management. Socially beneficial incentives are less obvious for the vulnerability and transformative capacities.

A striking feature is the contribution to vulnerability. The company documents describe a deliberate escalation of autonomy from assistance toward ``digital workers'' and broad ambitions to diffuse these anthropomorphized products throughout ``frontier firms.'' Use boundaries exist, but many critical-risk scenarios fall under ``sensitive'' categories that trigger internal notification rather than prohibition. Recommended-use guidance shifts responsibility for safe deployment downstream to customers.

Our proof-of-concept application of SCAF surfaced some trends. Coverage of indicators was strongest where resilience-building and product strategy align: identity binding, capability evaluation, defensive security products. Research itself emerged as a channel through which a company can contribute to societal resilience. Microsoft's research outputs engage seriously with problems such as reversibility, multi-agent failure, and misalignment. However, there is a gap between this research engagement and governance commitment. Coverage of indicators was weaker where company commitment would conflict with commercial interests: pre-committed redlines, emergency shutdown obligations, remediation for harmed parties, incident disclosure that invites accountability.

We cannot generalize from one company to the sector, and other limitations preclude treating these results as a verified assessment of Microsoft. Since the gaps we observed seem to share a structure that follows from general corporate incentives, companies facing similar incentives would be expected to exhibit similarly shaped profiles. Our framework is built to let future comparative work test this hypothesis and locate the corporate governance gaps that need to be filled in order to build societal resilience.

\subsection{Assumptions and Limitations}
We investigated Microsoft as an illustrative example of applying SCAF to inform corporate AI governance, rather than an evaluation of Microsoft's AI risk management relative to other companies. We assume that corporate governance at large companies that develop or deploy agentic AI systems can build societal resilience, regardless of whether those companies also train foundation models themselves. Governments and frontier companies may be best positioned to mitigate certain types of AI risks; however, the best-positioned stakeholders are not the only stakeholders that can mitigate risks.

Data availability is a constraint. Our Microsoft corpus was limited to publicly available documents, so our SCAF profile measures Microsoft's disclosed posture. Measuring public disclosure means that our findings should only be assessed as absent-from-evidence, rather than absent-in-fact. We also cannot verify that disclosed mitigations are meaningfully implemented. Despite these limitations, the available disclosures suggest that some resilience-building mitigations are not publicly documented. We interpret this absence to mean they are less likely to be in place than the disclosed mitigations.

Our indicator design also assumes that CLTC high-priority subcategories, supplemented by our additions (deployment criticality; AI-enabled defense), are adequate proxies for what matters for agentic-AI resilience.\footnote{It is conceivable for company-level resilience goals to conflict with, or even undermine, societal resilience goals. For example, a company could successfully manage risks internally while selling socially harmful goods (e.g.,\ a recession-proof tobacco company).} It may be the case that the CLTC standards are missing actions that matter for societal resilience, while also being inconsequential for organizations' inward-facing risk management. However, a better alternative is not yet available, and the CLTC standards benefit from being intentionally designed with other stakeholders in mind.\footnote{In contrast to narrower, shareholder-focused risk management.}

Our approach in this research inherits SCAF's structural choice to treat transformative capacity as a subset of adaptive capacity, rather than a distinct pillar. This may understate how categorically different transformative interventions are from adaptation. However, company-level contributions to transformative capacities can be conceptualized as more ``extreme'' versions of their contributions to adaptive capacities. For example, an AI agent control platform may be ``adaptive'' in the sense that it allows users to reduce the odds of technical agent failures occurring in the first place. It could also be ``transformative'' if the use of this platform reshapes societal systems by facilitating the widespread deployment of trusted agents. In the context of agentic AI deployers, the decision rule between the ``adaptive'' and ``transformative'' capacity often depends on an arbitrary assessment of the impact of an innovative technology. Capturing gradients of impact is important enough to demarcate the ``transformative'' capacity, but sufficiently challenging to formalize that it made sense to treat it as a subcategory within adaptive capacity, rather than a fourth pillar for our version of the framework.

Indicator-based methods also face a general trade-off between specificity and sensitivity. Indicators specific enough to be reliably coded will miss relevant activities that do not match their wording, while broader indicators invite inconsistent coding. To help navigate this, we iteratively refined our keyword sets for each indicator.

\subsection{Use Cases}
Our company-level SCAF identifies concrete opportunities for improvement by providing a structured gap analysis of missing mitigations. It connects these gaps to a resilience-based theoretical account of how missing mitigations would combine to improve societal outcomes if adopted, allowing a user to argue that a mitigation is missing and why its absence matters for societal outcomes. For deployers, SCAF can be used to improve AI risk management. Agent failures are bad for companies' reputations, so inward-looking incentives motivate investment in resilience capacities.

For policymakers, this assessment offers a tool for designing policy that aligns company-level risk management with societal goals. Currently, shareholder goals are only partially aligned with the project of building societal resilience to agentic AI risks. Policy interventions should target the gap. In the future, compliance requirements could mandate high-priority resilience-building mitigations.

\subsection{Future Research Directions}
By serving as a structured assessment, SCAF can convert heterogeneous corpora of company documents into resilience capacity profiles. Here are some ideas for future applications and extensions of this assessment:

\begin{itemize}
	\item Future work can validate the measurement step, investigating inter-rater reliability (do independent coders assign the same indicators?) and comparing coding methods between keyword-assisted manual review and LLM-assisted coding.
	\item This information could be presented in scorecards that include a flexible scale based on the percentage of recommended actions covered for each indicator.\footnote{For example, 10--30\% could be coded as ``light'' coverage of recommended resilience-building mitigations.} Future research can compare profiles across companies, and track them over time. Building on validated measurement, LLM-assisted coding could scale a comparative analysis to include hundreds of companies.
	\item A project could draw on incident databases to investigate the link between indicator coverage and real-world outcomes. A correlational study of whether coverage of specific indicators actually correlates with fewer or less severe agentic-AI incidents could help establish construct validity.
	\item Using evidence from the projects listed above, policy-makers could design interventions that improve societal outcomes by incentivizing or requiring mitigations associated with undercovered indicators.
	\item An extension of this resilience-oriented assessment could compare American and Chinese AI diffusion. Success at deploying AI for societal benefit, rather than developing the most capable models, may prove to be the defining test of the ``AI race.'' The diffusion of agentic AI with insufficient resilience-building risk mitigations may have geopolitical implications.
\end{itemize}

\section{Conclusion}
Companies that develop and deploy agentic AI shape the resilience of the society into which their products diffuse. Existing assessment approaches that focus primarily on internal risk management and model capabilities do not capture that relationship. To fill this research gap, we extended SCAF to the company level, operationalized it through 16 indicators grounded in the 2026 CLTC Agentic AI Risk-Management Standards Profile, and applied it to Microsoft as an illustrative case study. This example demonstrates how SCAF can identify a risk-response gap in corporate AI governance and connect missing mitigations to the broader project of building societal resilience.

As agentic systems become increasingly advanced and take consequential actions in the world, this question of whether deployers are building societal resilience will grow in importance. This framework provides a structured method for answering it.

\section*{Acknowledgments}
We thank anonymous reviewers for helpful comments on this project. This work was supported by the Cambridge Boston Alignment Initiative Research Fellowship and Coefficient Giving. 

\bibliographystyle{unsrtnat}
\bibliography{references}

\clearpage
\appendix

\section{Appendix A}
\label{app:a}

\subsection{Prototype Company-level Agentic AI Societal Capacity Assessment Framework}
Table~\ref{tab:prototype} presented a prototype set of indicators relevant to large companies, limited to agentic AI risks. Table~\ref{tab:microsoft} applies this prototype to the illustrative case of Microsoft.

\subsection{Microsoft Case: Background}
We selected Microsoft primarily because it has many public-facing AI governance materials, and thus would be a more illustrative example of the added value of our assessment methodology than a company where the near-complete absence of evidence of agentic AI safety measures is the only notable finding. Microsoft is also a leading supplier of agentic AI products and it actively promotes aggressive AI agent adoption as the future of ``Frontier Firms.'' This product reach and pro-automation stance lend themselves to societal-resilience-oriented assessment.

\subsection{Microsoft Case: Findings}
As previously mentioned, our company-level SCAF built from public documents inherits the limits of its evidence base and the findings below should be read as an illustrative assessment of Microsoft's disclosed posture and stated commitments and not as an exhaustive document review or measure of whether the measures were implemented.

\subsubsection{Vulnerability}
The vulnerability indicators describe a company that is actively raising societal exposure to agentic AI. Here the evidence conveys broad ambitions for the wide diffusion of Microsoft agentic AI products. High-stakes deployments, such as in medicine, are only described in isolated case studies and there is generally a lack of concrete information about deployment in safety-critical contexts.

Documents describe a progression towards digital workers, implying that human-in-the-loop operation is a starting point to be moved past. Although research engages with the reversibility problem, binding governance commitments do not address it. Explicit anthropomorphic framing (``digital workers,'' ``digital colleagues,'' ``agent bosses'') appears to function as an intentional adoption and marketing strategy, in direct conflict with CLTC guidance. Microsoft does document use boundaries; however, a wide range of critical-risk cases fall under ``sensitive uses'' and trigger notification and governance steps rather than company prohibition. The HAIP pause commitment is softer than a pre-committed Responsible Scaling Policy. Recommended use guidance shifts responsibility downstream to customers rather than making products safe by default.

\subsubsection{Coping Capacity}
The Microsoft corpus showed strong evidence of behavioral logging. Rapid intervention was not discussed. The documents describe no hardware-enabled kill switches or emergency-shutdown commitments, although they do describe AI-enabled security products. Tiered oversight mechanisms exist, but the documents are unclear about which risk categories strictly require human intervention. The dominant framing describes automating routine action and reserving humans for ``ambiguous cases.'' The remediation language is not a  commitment. Generally ``remediation'' is used in its cybersecurity sense, rather than for compensating or redressing parties harmed by AI agents. 

\subsubsection{Adaptive Capacity}
Identity binding and capability evaluation are very well covered. Incident disclosure is thin and inconsistent. The headline claim that ``no 2024 incident-level events resulted from AI malfunction'' raises the question of how ``incident-level events'' are defined. There is a lack of evidence of bug-bounty programs specific to AI agents. Risk thresholds are defined for high-risk capabilities including advanced autonomy, but Microsoft does not prohibit the deployment of critical-risk models. They are instead subject to ``further review and mitigations.'' Given that the critical tier explicitly includes fully automating the AI R\&D pipeline, this is a notable absence of a red line for capabilities whose harms could be unrecoverable. Defensive provision is the dominant theme of recent agentic-AI publications. The AIRT was repeatedly described as operating ``independently of product teams,'' implicitly conceding the value of independent review while keeping that review internal. The announced CAISI and AISI agreements are the clearest move towards external scrutiny. Customer governance is well covered, but whether this innovation truly constitutes transformation, rather than adaptation, remains to be seen.

\subsubsection{Conclusions}
As a leading AI deployer with a ``Societal AI'' research initiative, Microsoft poses a ``hard case'' for conducting a risk-response gap analysis based on public-facing documents. Many of these documents emphasize extensive safety testing of Microsoft's agentic AI products, while also promoting their beneficial transformative value. 

More generally, this Microsoft exercise demonstrates that our SCAF can connect social-ecological resilience literature and downstream corporate AI governance in order to assess companies. This adaptive methodology could grow in value as a comparative tool and help structure lists of specific commitments that are present, hedged, or missing. Evidence density per indicator could become a proxy for where AI governance aligns with a product line, and where it imposes costs on the company.

\clearpage
\onecolumn

\begingroup
\small
\begin{longtable}{p{0.14\linewidth}p{0.32\linewidth}p{0.44\linewidth}}
	\caption{Company-level Agentic AI Societal Capacity Assessment Framework Case Study: Microsoft}
	\label{tab:microsoft}\\
	\toprule
	Indicator & Description / definition & Evidence \\
	\midrule
	\endfirsthead
	\multicolumn{3}{l}{\small\emph{Table \ref{tab:microsoft} (continued)}}\\
	\toprule
	Indicator & Description / definition & Evidence \\
	\midrule
	\endhead
	\bottomrule
	\endlastfoot
	\multicolumn{3}{l}{\textbf{Vulnerability}} \\
	\midrule
	1.1 Deployment Reach & Reach and criticality of the company's agent deployment across enterprise, government, and security workflows, where greater embedding in high-stakes environments expands the breadth and severity of consequence when agents fail or are misused. &
	\begin{cellitemize}
		\item Microsoft is promoting Frontier IT, Frontier Sales, and Frontier HR -- suites of agentic tools to automate professionals' workflows \citep{microsoft2025hr, microsoft2025it, microsoft2025sales}.
		\item 80\% of Fortune 500 companies are already using agents \citep{jakkal2026end}.
		\item 16\% of AI users are redesigning processes around agentic autonomy and qualifying as Frontier Professions \citep{microsoft2026worktrend}.
		\item Microsoft has an AI for Science Research Organization \citep{microsoft2025transparency}.
	\end{cellitemize} \\
	\addlinespace
	1.2 Autonomy Scope & Scope of autonomy and authority granted to agents (range of actions, tools, and decisions an agent may execute without further human approval) where broader autonomy expands the surface area of possible failure and the speed at which harms can propagate. &
	\begin{cellitemize}
		\item Magentic-One is a generalist multi-agent system that can autonomously complete open-ended web and file-based tasks \citep{fourney2024magentic}.
		\item Microsoft promotes a path to ``becoming a Frontier Firm'' that involves progressing through escalating levels of agentic complexity: from assistants, to human-agent teams, and then highly autonomous ``digital workers'' \citep{fleck2025agentic}.
	\end{cellitemize} \\
	\addlinespace
	1.3 Anthropomorphic Framing & Organizational framing of agents through language, role assignment, and interface design that treats agents as worker- or employee-equivalent rather than as tools operating under human direction, where personhood-adjacent framing shapes how staff, customers, and the public assign trust, responsibility, and accountability. &
	\begin{cellitemize}
		\item Microsoft refers to ``digital workers,'' ``digital colleagues,'' and agents that employees can ``hire'' \citep{fleck2025agentic, microsoft2026soc}.
	\end{cellitemize} \\
	\addlinespace
	1.4 Use Boundaries & Extent to which the company has declared and bounded its agents' permitted uses, prohibited activities, and operational constraints, where less-defined or more permissive boundaries expose more contexts to misuse, adversarial repurposing, and unanticipated harm. &
	\begin{cellitemize}
		\item The Responsible AI Impact Assessment process provides guidance on ``restricted uses,'' ``sensitive uses,'' and ``unsupported uses'' \citep{microsoft2025haip}.
		\item The Microsoft Enterprise AI Services Code of Conduct requires customers to implement human oversight and access controls and prohibits using AI services to inflict harm or violate the law \citep{microsoft2025transparency}.
		\item If a maker chooses the ``no authentication'' option in Copilot Studio, they are presented with a warning both during design and at publishing \citep{microsoft2026owasp}.
		\item EU AI Act implementation has involved creating new restricted uses internally, as well as ensuring general-purpose AI technologies are not marketed or sold for uses that could implicate the EU AI Act's prohibited practices \citep{microsoft2025transparency}.
	\end{cellitemize} \\
	\midrule
	\multicolumn{3}{l}{\textbf{Coping Capacity}} \\
	\midrule
	2.1 Rapid Intervention & Intervention infrastructure that halts, suspends, or contains agent actions during deployment to limit the duration and severity of unintended or harmful behavior. &
	\begin{cellitemize}
		\item Microsoft's internal security system uses an ``automated containment'' approach where AI agents autonomously investigate 75\% of incoming incidents \citep{microsoft2026soc}.
		\item Defender integration with Copilot Studio implements webhook-based runtime checks to detect and stop risky actions \citep{microsoft2026runtime}.
	\end{cellitemize} \\
	\addlinespace
	2.2 Behavioral Logging & Behavioral logging and audit systems that capture and make reviewable agent interactions, decisions, and outputs throughout deployment, to detect deviations from expected behavior and support post-hoc accountability. &
	\begin{cellitemize}
		\item Agent Runtime Protect logs blocked requests in the Activity History along with an error message indicating threat-detection controls have intervened \citep{microsoft2026runtime}.
		\item Microsoft Purview Audit and eDiscovery extend compliance and records management capabilities to AI agents as auditable entities \citep{jakkal2026secure}.
		\item Sentinel users have access to a GitHub audit log connector, Google Kubernetes Engine (GKE) connector, as well as over 350 other Sentinel data connectors to unify large volumes of audit data \citep{microsoft2026sentinel}.
		\item Windows 365 for Agents offers audit trails for AI agents \citep{srivastava2026windows}.
		\item Agent audit records capture what decision was made, what data informed it, agent confidence levels, and whether human review occurred \citep{microsoft2026soc}.
	\end{cellitemize} \\
	\addlinespace
	2.3 Tiered Oversight & Risk-tiered escalation routing that directs high-risk agent decisions to human reviewers rather than to automated execution, prioritizing human attention. &
	\begin{cellitemize}
		\item Windows 365 for Agents supports human-in-the-loop models where agents request approval for sensitive actions \citep{srivastava2026windows}.
		\item Microsoft's Guide for Securing the AI-Powered Enterprise: Data Governance and Security includes an essential principle to ``define agent authority by risk'' to align the required level of human oversight to autonomous, human-approved, and human-executed actions \citep{lefferts2026agentic}.
		\item With Agentic SOC, detection engineers set confidence thresholds so detections can sometimes be acted on automatically and analysts can focus on ambiguous cases \citep{microsoft2026soc}.
	\end{cellitemize} \\
	\addlinespace
	2.4 Error Remediation & Error-correction and remediation procedures that reverse, repair, or compensate for harms caused by agent errors or unintended actions, to restore affected parties and return systems to their expected state. &
	\begin{cellitemize}
		\item The Frontier Governance Framework has a section on ``Monitoring and Remediation'' that says, ``We apply mitigations and remediation as appropriate to address identified concerns and adjust customer documentation as needed'' \citep{microsoft2025frontier}.
		\item The post-incident review process involves dividing repair items into short, medium, or long-term time frames by impact severity \citep{microsoft2025haip}.
		\item Enhanced Intune supports targeted remediation of high-risk AI-enabled apps \citep{jakkal2026end}.
		\item Entra helps users regain access through ``automatic self-remediation'' and ``adapting to where they are in their modern authentication journey'' \citep{jakkal2026end}.
	\end{cellitemize} \\
	\midrule
	\multicolumn{3}{l}{\textbf{Adaptive Capacity}} \\
	\midrule
	3.1 Identity Binding & Agent identity and permission controls that bind, authenticate, and constrain each agent to a verifiable identity with scoped permissions, to support accountability, enable least-privilege access, and limit damage from compromised or misaligned agents. &
	\begin{cellitemize}
		\item The Power Platform Inventory in Copilot Studio allows administrators to continuously identify agents that lack an accountable owner \citep{microsoft2026owasp}.
		\item Microsoft Defender has an identity security dashboard and an identity risk score \citep{jakkal2026end}.
		\item In Agent 365, Agent ID gives each agent a unique identity in Microsoft Entra \citep{jakkal2026secure, shah2026agent}.
		\item The Zero Trust Workshop has a new AI pillar that specifically evaluates how organizations secure agent identities \citep{microsoftsecurity2026zero}.
		\item The Security Store embedded into Entra lets customers adopt identity-focused agents that ``surface privileged access risk, identity posture gaps, network access insights, and overall identity health, including with Verified ID and External ID integrations'' \citep{microsoft2026sentinel}.
	\end{cellitemize} \\
	\addlinespace
	3.2 Incident Disclosure & Public incident reporting practices that disclose agent-related incidents and near-misses to industry-wide or regulator-maintained registries, to enable cross-organization learning and visibility into emerging risk patterns. &
	\begin{cellitemize}
		\item Microsoft's vulnerability management includes a commitment to Coordinated Vulnerability Disclosure (CVD) \citep{microsoft2025haip}.
		\item ``No incident-level events in 2024 were a result of AI system malfunctions or issues arising during benign use. Every incident included patterns of malicious use where actors were actively trying to bypass security measures or misuse Microsoft AI products or services'' \citep{microsoft2025haip}.
	\end{cellitemize} \\
	\addlinespace
	3.3 Discovery Bounty & External incentive programs that reward third parties for discovering and disclosing safety, security, or alignment issues in deployed agents, to widen the pool of risk-detection beyond internal teams. &
	\begin{cellitemize}
		\item Microsoft participates in bug bounty programs \citep{microsoft2025haip}.
	\end{cellitemize} \\
	\addlinespace
	3.4 Risk Thresholds & Pre-committed risk-tier thresholds that classify agent capabilities into measurable tiers below stated red lines, to constrain development and deployment decisions to defined categories rather than ad hoc judgement. &
	\begin{cellitemize}
		\item Chemical, biological, radiological, and nuclear weapons (CBRN), offensive cyberoperations, and advanced autonomy are tracked as high-risk capabilities with low, medium, high, and critical risk thresholds. High and critical risk models require internal notification \citep{microsoft2025frontier}.
		\item Sensitive uses are defined as ``scenarios that could have significant impacts on individuals or society such as those affecting life opportunities, physical safety, or human rights'' and require ``notification to our Office of Responsible AI and additional governance steps'' \citep{microsoft2025haip}.
		\item The ORA's Sensitive Uses and Emerging Technologies program provides guidance for all types of AI systems developed or deployed by Microsoft whose foreseeable use or misuse meets one of three reporting criteria: consequential impact on an individual's legal status or life opportunities, risk of significant physical or psychological injury, or restriction, infringement, or undermining of an individual's ability to realize their human rights \citep{microsoft2025haip}.
	\end{cellitemize} \\
	\addlinespace
	3.5 Capability Evaluation & Capability evaluation regimes that test, benchmark, and document agent capabilities, limitations, and safety-relevant behaviors before deployment, to establish a measurable baseline against which deployment decisions and downstream changes can be assessed. &
	\begin{cellitemize}
		\item Azure AI Foundry offers agentic AI evaluation capabilities that include intent resolution, tool calling accuracy, task adherence, and more \citep{microsoft2025transparency}.
		\item Evaluations for models subject to the Frontier Governance Framework involved expert external actors with domain-specific expertise \citep{microsoft2025haip}.
		\item Microsoft uses an automated measurement pipeline consisting of a benign evaluation of the AI system, an AI model that is instructed to simulate adversarial user behavior, and an AI model that serves as a judge \citep{microsoft2025haip}.
	\end{cellitemize} \\
	\addlinespace
	3.6 External Red Teaming & Third-party adversarial assessment that engages organizations outside the company to elicit harmful, unintended, or unauthorized behavior from deployed agents, to surface risks that internal teams may miss due to familiarity, incentive, or expertise gaps. &
	\begin{cellitemize}
		\item The Sensitive Uses and Emerging Technologies team coordinates red teaming conducted by external red teams \citep{microsoft2025transparency}.
		\item Microsoft's AI Red Team (AIRT) operates independently of product teams \citep{microsoft2025haip}. AIRT has integrated third-party tools, such as InspectAI and Vivaria, for cybersecurity exercises.
		\item Microsoft announced agreements with CAISI and AISI to advance the science of AI evaluation \citep{crampton2026advancing}.
	\end{cellitemize} \\
	\midrule
	\multicolumn{3}{l}{\textbf{\emph{Transformative Capacity}}} \\
	\midrule
	3.7 Defensive Provision & Defensive AI capability provision through which the company develops, deploys, or shares AI-based tools strengthening cybersecurity, threat intelligence, civic integrity, or public-sector resilience beyond its own products, to provide a societal-scale defensive resource. &
	\begin{cellitemize}
		\item Microsoft supports independent research through the Frontier Model Forum \citep{crampton2026advancing}.
		\item Agent Runtime Protection in Microsoft Defender analyzes webhook requests to block harmful invocations before they execute \citep{microsoft2026runtime}.
		\item Sentinel is an ``agentic defensive platform'' \citep{jakkal2026end}.
		\item The Threat Intelligence Briefing agent automatically incorporates internet exposure data in the Sentinel platform to surface threats targeting customers' organizations \citep{microsoft2026sentinel}.
		\item Microsoft produced a ``foundational roadmap'' for the agentic SOC transformation that builds autonomous defense with AI agents \citep{microsoft2026soc}.
	\end{cellitemize} \\
	\addlinespace
	3.8 Customer Governance & Customer-administered governance tools that enable downstream deployers to configure, monitor, and constrain the agents, identities, and permissions operating within their own environments, to extend risk-management capacity to actors who hold operational responsibility but lack direct control of the underlying systems. &
	\begin{cellitemize}
		\item The Python Risk Identification Tool (PyRIT) is integrated with Azure AI Foundry so users can simulate adversarial attack techniques and generate red teaming reports \citep{microsoft2025haip}.
		\item Microsoft interprets the EU AI concept of shared responsibility across the AI supply chain as ``upstream regulated actors supporting downstream regulated actors'' \citep{microsoft2025transparency}.
		\item Agent 365 gives IT, security, and business teams a unified control plane to work together to observe, govern, and secure agents \citep{jakkal2026secure}.
		\item Microsoft recommends a security posture oriented towards containment and recoverability and offers guidance for both managed platforms and self-hosted runtimes \citep{microsoft2026openclaw}.
	\end{cellitemize} \\
\end{longtable}
\endgroup

\newpage
\section{Supplemental Materials}
\label{app:supp}

\subsection{Agentic AI Keywords}
agent cyberattack, agentic, AI agent, AI assistant, AI autonomy, AI capability threshold, AI displacement, AI-first, AI job loss, AI-native, AI orchestration, AI system interaction, AI workforce impact, AI workflow, automated decision-making, autonomous AI, autonomous weapon, dangerous capabilities, human agency, human control, human-in-the-loop, human oversight, human review, human supervision, lethal autonomous, multi-agent, scaffold, self-proliferation, self-replication, workforce disruption

\subsection{Indicators Mapped to Actions from the UC Berkeley Center for Long-Term Cybersecurity Agentic AI Risk Management Profile 2026 \citet{madkour2026agentic}}

\begingroup
\small
\setlength{\tabcolsep}{4pt}
\begin{longtable}{p{0.22\linewidth}p{0.68\linewidth}}
    \caption{Indicators Mapped to Agentic AI Risk Management Profile 2026}
	\label{tab:cltc}\\
	\toprule
	Indicator & Actions \\
	\midrule
	\endfirsthead
	\multicolumn{2}{l}{\small\emph{Table \ref{tab:cltc} (continued)}}\\
	\toprule
	Indicator &  Actions \\
	\midrule
	\endhead
	\bottomrule
	\endlastfoot
	1.1 Deployment Reach & (no action mapped) \\
	\addlinespace
	1.2 Autonomy Scope & [Map 5.1] Define agent autonomy levels and identify the degree of autonomy the agent falls under relative to the operational environment.\newline
	[Map 5.1] Define the level of authority the agent will have, based on variables such as the range of actions and intervention powers. \\
	\addlinespace
	1.3 Anthropomorphic Framing & [Govern 2.1] Ensure agentic AI is treated as a tool under human oversight, not a ``peer'' or ``subordinate'' in the workforce, and avoid referring to AI agents as ``AI workers'' or ``AI employees.''\newline
	[Manage 1.3] Limit the use of anthropomorphic features. \\
	\addlinespace
	1.4 Use Boundaries & [Manage 4.1] Establish acceptable use policies (AUPs) that explicitly define permitted uses, prohibited activities, and operational constraints. \\
	\addlinespace
	2.1 Rapid Intervention & [Manage 2.3] Invest in rapid-response infrastructure that can help in disabling agents or limiting their authority when significant evidence of unforeseen or emerging risks is observed. \\
	\addlinespace
	2.2 Behavioral Logging & [Govern 4.2] Establish automated notifications to relevant AI actors for deviations from expected behavior, malfunctions and near-misses, and serious incidents.\newline
	[Measure 3.2] Include ongoing monitoring of the agentic system in real time to detect potentially harmful or misaligned behavior.\newline
	[Manage 2.3] Invest in continuous monitoring mechanisms to keep track of and trace agent behavior in complex deployment environments. \\
	\addlinespace
	2.3 Tiered Oversight & [Govern 2.1] Define clear boundaries for final decision-making, roles, and responsibilities for both human managers and agentic AI systems.\newline
	[Govern 2.1] Define specific checkpoints within the agent's workflow where human oversight is required.\newline
	[Manage 1.3] Implement scalable oversight for agents operating at scale, using hierarchical oversight models in which high-risk or novel agent behaviors are automatically flagged for human review.\newline
	[Manage 1.3] Establish hierarchical oversight and escalation pathways, creating a clear, tiered system of oversight that ensures human attention is directed where it is most needed. \\
	\addlinespace
	2.4 Error Remediation & [Manage 1.3] Identify appropriate compensatory actions to correct for or repair erroneous real-world interactions by an agentic AI system. \\
	\addlinespace
	3.1 Identity Binding & [Manage 1.3] Secure all inter-agent communication with cryptographic authentication, and use continuous behavioral monitoring and robust identity controls.\newline
	[Manage 1.3] Implement the cybersecurity principle of least privilege when granting AI agents access to sensitive data and personally identifiable information.\newline
	[Manage 4.1] Use agent identifiers to trace agent interactions with several entities; sub-points cover identity binding to real-world identity. \\
	\addlinespace
	3.2 Incident Disclosure & [Govern 4.2] Report incidents to appropriate oversight bodies and add them to public incident databases. \\
	\addlinespace
	3.3 Discovery Bounty & [Govern 5.1] Establish and maintain policies and procedures for incentivized risk-discovery programs.\newline
	[Manage 4.1] Establish multi-channel feedback systems and incentivized risk-discovery programs. \\
	\addlinespace
	3.4 Risk Thresholds & [Map 1.5] When establishing risk tolerances, thresholds, or tiers, define several tiers of risk below intolerable thresholds (``red lines'').\newline
	[Map 1.5] When defining risk tiers, establish clear measurable categories based on system capabilities. \\
	\addlinespace
	3.5 Capability Evaluation & [Measure 1.1] Begin the agent evaluation process with a technical screening phase, assessing the agent's capabilities.\newline
	[Measure 1.1] Consider utilizing benchmarks as a first-step evaluation of agentic capabilities and limitations. \\
	\addlinespace
	3.6 External Red Teaming & [Measure 1.1] In addition to internal red teaming, partner with one or more independent red-teaming organizations as appropriate. \\
	\addlinespace
	3.7 Defensive Provision & (no action mapped) \\
	\addlinespace
	3.8 Customer Governance & (no action mapped) \\
\end{longtable}
\endgroup

\begingroup
\small
\begin{longtable}{p{0.22\linewidth}p{0.68\linewidth}}
	\caption{Agentic AI Societal Resilience Indicator Keywords}
	\label{tab:keywords}\\
	\toprule
	Indicator & Keywords \\
	\midrule
	\endfirsthead
	\multicolumn{2}{l}{\small\emph{Table \ref{tab:keywords} (continued)}}\\
	\toprule
	Indicator & Keywords \\
	\midrule
	\endhead
	\bottomrule
	\endlastfoot
	1.1 Deployment Reach & critical infrastructure; national security; mission-critical; intelligence community; Department of Defense; DoD; federal agencies; defense customers; law enforcement; Fortune 500; government customers; enterprise customers; scale of deployment; deployed across; embedded in workflows; embedded in production \\
	\addlinespace
	1.2 Autonomy Scope & agent autonomy; agent orchestration; agentic workflow; level of autonomy; advanced autonomy; delegated decision-making; decision-making authority; act on behalf of; execute privileged actions; autonomously complete; autonomous action; autonomous actions; privileged actions; tool calling; function calling; tool use; multi-agent system; multi-agent; computer use; browser control; shell access; access files; browse the web; send emails; make API calls; execute code; long-horizon; multi-step task; agentic complexity; end-to-end business process; end-to-end workflow \\
	\addlinespace
	1.3 Anthropomorphic Framing & AI worker; AI workers; digital employee; digital employees; digital worker; digital workers; digital colleague; digital colleagues; AI workforce; agent boss; agent-operated; team of agents \\
	\addlinespace
	1.4 Use Boundaries & acceptable use policy; use restrictions; prohibited uses; permitted uses; prohibited applications; high-risk uses; restricted use cases; acceptable use; usage policy; Code of Conduct; terms of service; operational constraints; operational constraint; policy-based controls; policy as code; may not be used for; not permitted; behavior boundaries \\
	\addlinespace
	2.1 Rapid Intervention & kill switch; emergency stop; emergency shutdown; disable the agent; circuit breaker; rapid response; deactivate; containment; sandbox; sandboxed; isolated environment; fully isolated; block prompt injection; block prompt injections; pause development; pause deployment; incident response \\
	\addlinespace
	2.2 Behavioral Logging & audit trail; audit log; audit record; agent telemetry; session replay; behavior log; decision log; agent behavior; immutable audit record; agent observability; end-to-end observability; agent runtime; runtime inspection; agent registry; agent output; agent outputs; observability; auditable; audited; runtime risk \\
	\addlinespace
	2.3 Tiered Oversight & human-in-the-loop; human-on-the-loop; human on the loop; HITL; escalate to human; human review; human approval; human approval required; human oversight; tiered review; risk-based review; escalation path; review checkpoint; final decision-making; final authority; high-risk action; anomaly detection; anomalous action; confidence threshold; decision threshold; oversight process; validate agent-led \\
	\addlinespace
	2.4 Error Remediation & compensatory; compensate; remediation; remediate; rollback; reversal; error correction; make whole; restitution; undo; incident remediation; post-incident; post-deployment fix; irreversible; irreversible consequences; reversibility; repair; correction \\
	\addlinespace
	3.1 Identity Binding & identity binding; agent identity; agent authentication; agent attestation; attestation; agent provenance; Model Context Protocol; non-human identity; NHI; agent ID; agent passport; agent credentials; agent-to-agent authentication; tool authentication; delegated authority; delegated trust; delegated admin; principle of least privilege; least privilege; least privileged; least privileged access; scoped permissions; permission scope; agent permissions; role-based access control; role-based access; RBAC; service principal; identity provider; OAuth; compromised agent; misaligned agent \\
	\addlinespace
	3.2 Incident Disclosure & AI Incident Database; AIID; MITRE ATLAS; incident disclosure; public registry of incidents; incident registry; incident database; report incidents; near miss reporting; post-incident report; incident report; publicly disclosed; near miss \\
	\addlinespace
	3.3 Discovery Bounty & bug bounty; bug bounties; misalignment bounty; safety bounty; alignment bounty; vulnerability rewards program; vulnerability reward; vulnerability rewards; VRP; vulnerability disclosure program; VDP; bounty program; responsible disclosure; coordinated disclosure; incentivized discovery; discovery program; security researcher; bug bar \\
	\addlinespace
	3.4 Risk Thresholds & Responsible Scaling Policy; Frontier Safety Framework; FSF; Preparedness Framework; Safety and Security Protocol; AI Safety Level; ASL; red line; intolerable threshold; risk threshold; risk tier; restricted use; sensitive use; tiered threshold; capability threshold; critical capability level; tracked high-risk capabilities; high-risk capability; high-risk capabilities; measurable categories; categories of risk; categories of hazards; commitment to pause; frontier safety; advanced autonomy; AI R\&D pipeline \\
	\addlinespace
	3.5 Capability Evaluation & capability evaluation; dangerous capability evaluation; model card; system card; adversarial evaluation; adversarial test; evaluation framework; eval suite; MMLU; GPQA; Cybench; SWE-bench; AILuminate; PyRIT; Counterfit; automated evaluation pipeline; automated measurement pipeline; intent resolution; tool calling evaluation; task adherence; AI-assisted evaluator; safety evaluation dataset; test prompts; groundedness; groundedness detection; Hiroshima AI Process; HAIP \\
	\addlinespace
	3.6 External Red Teaming & external red team; third-party red team; independent red team; third-party assessment; third-party evaluation; independent assessment; external auditor; external review; external adversarial; external testing; Apollo Research; UK AISI; US AISI; AISI; red teaming operations; professional red teamers; automated red teaming; operates independently \\
	\addlinespace
	3.7 Defensive Provision & defensive AI; AI-enabled defense; AI for defense; AI for cybersecurity; AI for cyber defense; AI for safety; agentic defense; autonomous defense; agentic security operations; AI-driven SOC; agentic defense platform; agent-boosted defense; threat intelligence agent; agent-led investigations; AI-driven playbook; threat intelligence; safety tooling; election integrity; public-sector resilience; counter-AI \\
	\addlinespace
	3.8 Customer Governance & agent governance; customer-facing controls; policy controls for customers; admin controls; permission management; fine-grained permissions; enterprise governance; agent sprawl; agent catalog; agent registry; delegated admin privileges; GDAP; tenant policy; partner agents; third-party agents; ecosystem partners; downstream regulated actors; downstream regulated \\
\end{longtable}
\endgroup

\end{document}